# Maintaining SUV Accuracy in Low-Count PET with PETfectior: A Deep Learning Denoising Solution

Yamila Rotstein Habarnau, Nicolás Bustos, Paola Corona, Christian González, Sonia Traverso, Federico Matorra, Francisco Funes, Juan Martín Giraut, Laura Pelegrina, Gabriel Bruno and Mauro Namías*

Fundación Centro Diagnóstico Nuclear (FCDN), Buenos Aires C1417CVE, Argentina

*Corresponding author: mnamias@fcdn.org.ar

## Abstract

**Background:** The diagnostic image quality of positron emission tomography (PET) acquisitions strongly depends on the administered radiotracer activity and the acquisition time. However, keeping these variables as low as possible is desired in order to reduce the patient radiation exposure and the radiopharmaceutical related costs of PET examinations. PETfectior is an artificial intelligence-based software that processes PET scans and increases the signal-to-noise ratio, obtaining high-quality images from low-count-rate images. In this study, we perform an initial clinical validation of PETfectior on images acquired with half of the counting statistics required to meet the most recent European Association of Nuclear Medicine (EANM) quantitative standards for $^{18}$F-FDG PET, evaluating lesions detectability, quantitative performance and image quality.

**Materials and methods:** 258 patients referred for $^{18}$F-FDG PET/CT were prospectively included. The standard-of-care scans (100% scans) were acquired and reconstructed according to EARL standards 2. Besides, half-counting-statistics versions of these images were obtained from list-mode data and processed with PETfecftior (50%+PETfectior scans). All the oncologic lesions were segmented on both versions of the PET/CT images either manually or automatically and lesions detectability was evaluated. The maximum SUV of the lesions was measured and the quantitative concordance of 50%+PETfectior and 100% images was evaluated. Subjective image quality was visually assessed by two experienced physicians.

**Results:** 1649 lesions were detected in a total of 198 studies. The 50%+PETfectior images showed high sensitivity for lesion detection (99.9%) and only one false positive was detected. Quantitative concordance of $SUV_{max}$ measured on 100% and 50%+PETfectior images was within ±12.5% (95% limits of agreement), with a bias of -1.01%. Image quality of the 50%+PETfectior images was rated equal to or better than the standard-of-care images.

**Conclusion:** PETfectior can safely be used in the clinical practice at half counting statistics, with high sensitivity and specificity, low quantitative bias and high subjective image quality.

## 1. Introduction

Fluorodeoxyglucose ($^{18}$F-FDG) positron emission tomography combined with computed tomography (PET/CT) is an imaging modality routinely used in oncologic practice for detection, staging and assessment of treatment response across a wide range of pathologies. The diagnostic image quality of PET acquisitions depends on the number of detected annihilation photons, which directly influences image signal-to-noise ratio (SNR) and lesion detectability. This is determined by several factors, among which the counting statistics is critical and directly proportional to the administered radiotracer activity and the acquisition time. In general, higher injected activity results in better image quality as long as the peak noise equivalent count rate is not exceeded. However, the administered radiotracer activity must be constrained by patient radiation exposure. Besides, both increased injected activities and longer acquisition times result in higher overall costs of PET examinations, due to greater radiotracer usage and reduced scanner throughput.

Accurate lesion uptake quantification is essential for reliable diagnosis, prognosis and treatment response assessment in PET imaging. The standardized uptake value (SUV) is widely used as a measure of $^{18}$F-FDG uptake and the maximum SUV within a lesion is commonly reported and monitored ($SUV_{max}$). However, quantitative values in PET depend on several factors, and calibration and harmonization strategies are needed to ensure adequate repeatability and reproducibility among studies.

Over the last decades, many improvements and new technologies have been introduced to PET scanners in order to enhance the SNR and improve image quality and lesion detectability, while keeping the radiation dose as low as possible. Such developments include the introduction of time-of-flight (TOF), point spread function (PSF) modelling during image reconstruction, the integration of digital photon detectors and, more recently, total-body PET systems.

Also, deep learning denoising and reconstruction techniques have been introduced to further enhance image quality and reduce noise (Hashimoto et al., 2024). Most of these denoising methods utilize paired low-dose and standard-dose images from specific scanner models to train their machine learning models. Also, hybrid deep learning + iterative reconstruction algorithms have recently been introduced in clinical systems (Liao et al., 2023). Deep-learning-based denoising technology has already been implemented in commercial PET scanners (Wang et al., 2022). A commercial software (SubtlePET, Subtle Medical, Menlo Park, CA, USA), based on the work by Xu et al. (2017) is also available (Bonardel et al., 2022; Chaudhari et al., 2021; Katsari et al., 2021; Weyts et al., 2022, 2023). The development of deep-learning techniques to recover high-quality imaging from low-dose scans remains an open and highly active area of research, as evidenced by the existence of dedicated challenges (UDPET Challenge 2024).

PETfectior is an image processing software aimed at increasing the signal-to-noise ratio in PET, obtaining high-quality images from low-count-rate images. It is based on deep convolutional neural networks that were trained through a proprietary method to distinguish signal from noise components. It was trained using images from only one PET scanner (GE Discovery 710). However, thanks to a proprietary domain transfer method, a fine-tuned model for each scanner and reconstruction parameters can be obtained. This technique is more efficient than collecting training images from different tomographs, that will never cover the entire universe of combinations of scanner models and reconstruction parameters.

In this article we describe the clinical validation of PETfectior on images acquired with half of the counting statistics required to meet the most recent European Association of Nuclear Medicine (EANM) quantitative standards (Accreditation Specifications – EANM EARL; Boellaard et al., 2015; Koopman et al., 2016). Unlike other commercial denoising software, PETfectior can also improve image quality of images acquired with full counting statistics. The evaluation of this application of PETfectior is outside the scope of this article.

The denoising performance was evaluated on PET/CT images from two different scanners at our institution. We compared the standard-of-care images currently used in our clinical practice and a half-counting-statistics version of these images processed by PETfectior. Lesion detectability and quantitative concordance of these two image versions were evaluated. Subjective image quality was also evaluated for a subset of images processed with PETfectior.

Lesion detectability was assessed by analyzing every lesion on each pair of PET/CT scans. Given the enormous effort required to manually segment these lesions, a deep-learning-based method aimed at the segmentation of oncologic lesions was used to ease the physician's workload.

## 2. Materials and methods

### 2.1. PETfectior software

PETfectior is software as a medical device (SaMD) which is provided as a service (SaaS) to process PET scans. It was developed at Fundación Centro Diagnóstico Nuclear (FCDN), Buenos Aires, Argentina. It uses proprietary technology to increase the signal-to-noise ratio of PET images using a 3D-UNET-based architecture (Çiçek et al., 2016; Ronneberger et al., 2015). It received regulatory clearance for commercial use in January 2025 by ANMAT, the national food, drugs and medical technology authority of Argentina.

### 2.2. Patient population

Patient images were acquired at FCDN. The data collection and analysis protocol was approved by the Institutional Internal Review Board and all patients signed informed consent. A prospective study was conducted between June 2022 and April 2025, during which oncological $^{18}$F-FDG PET/CT scans were acquired in list mode to derive half-counting-statistics versions of the standard-of-care scans. Patients were included randomly, representing the local population distribution and frequency of cancer types imaged with $^{18}$F-FDG.

We included a total of 258 (124 female and 134 male) $^{18}$F-FDG PET/CT oncologic studies acquired with 2 different scanner models: GE Discovery 610 and GE Discovery 710. The number of patients per scanner, as well as their ages and weights are summarized in Table 1. Patient age and weight distributions are also shown in Figure 1.

Two cohorts were analyzed separately. Cohort 1 consisted of 46 patients (18 female and 28 male) whose PET/CT were acquired on a GE Discovery 710 scanner. This cohort was manually analyzed by experienced physicians who labelled all the lesions and judged the image quality of the images.

Cohort 2 consisted of 212 studies (106 women, 106 men), whose PET/CT were acquired mostly on a GE Discovery 610 scanner to test the domain-transfer performance. These studies were randomly selected, although pediatric patients were specifically included in order to have enough samples to validate the method for this particular population. These studies were semi-automatically processed, as described below.

| Scanner | Sex | #Patients | Age [yr] | | | Weight [kg] | | |
|---|---|---|---|---|---|---|---|---|
| | | | Mean | Minimum | Maximum | Mean | Minimum | Maximum |
| GE Discovery 610 | Female | 99 | 49.8 | 9 | 83 | 66.8 | 38 | 121 |
| | Male | 104 | 48.3 | 5 | 90 | 70.1 | 19 | 114 |
| | Total | 203 | 49.0 | 5 | 90 | 68.5 | 19 | 121 |
| GE Discovery 710 | Female | 25 (18) | 44.9 (57.6) | 9 (31) | 83 (83) | 69.6 (75.3) | 31 (42) | 125 (125) |
| | Male | 30 (28) | 59.5 (62.8) | 9 (30) | 84 (84) | 80.2 (81.8) | 51 (61) | 122 (122) |
| | Total | 55 (46) | 52.8 (60.8) | 9 (30) | 84 (84) | 75.4 (79.2) | 31 (42) | 125 (125) |

**Table 1:** Population demographics for each scanner. Values shown in red and in parentheses correspond to cohort 1; values in black correspond to cohorts 1 and 2 altogether.

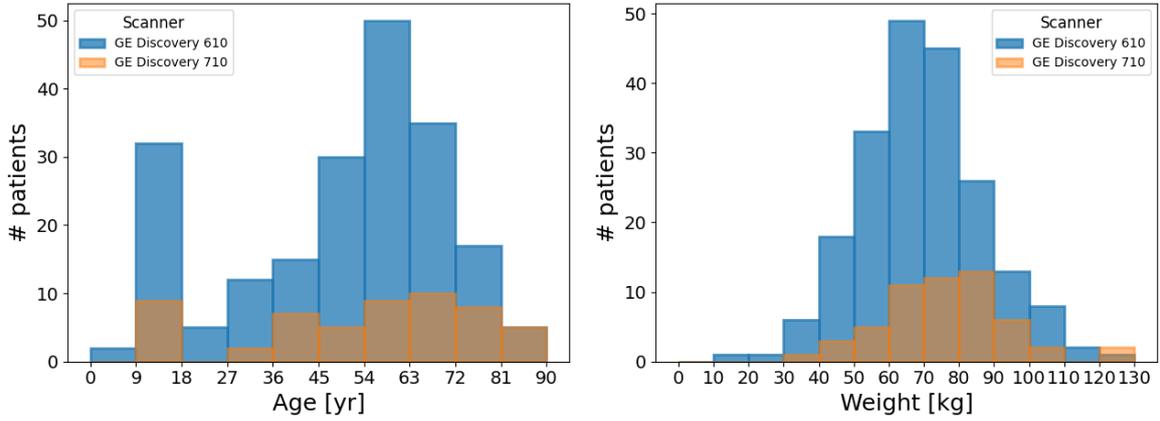

**Figure 1:** Patient age at the time of scan (left) and patient weight (right) distributions for the 258 studied patients.

### 2.3. PET/CT protocol

The reference PET image (with 100% counting statistics, from now on "100% image") was acquired and reconstructed according to EARL standards 2 (Accreditation Specifications – EANM EARL). The time-activity product scheme was already optimized as the minimum possible to comply with these specifications before introducing a further counting statistic reduction (Koopman et al., 2016). The acquisition and reconstruction protocols for each scanner are detailed in Table 2. Besides, a second image was reconstructed from list-mode data with half acquisition time and then processed with PETfectior (from now on, "50%+PETfectior image").

| Scanner | Overlap between bed positions | Acquisition time [seconds/bed] | Injected activity | Reconstruction algorithm | Voxel size (x, y, z) [mm] | Post-filters |
|---|---|---|---|---|---|---|
| GE Discovery 610 | 50% | 90 | $239\ MBq \times \left(\dfrac{w}{75\ kg}\right)^2$ | VPHD-S, 3 iterations, 16 subsets | (3.65, 3.65, 3.27) | X-Y: 4.4 mm FWHM Z: STD |
| GE Discovery 710 | | 120 | | VPFX-S, 2 iterations, 18 subsets | | X-Y: 5.2 mm FWHM, Z: STD |

**Table 2:** Acquisition and reconstruction protocols for each scanner. Here, $w$ stands for the patient's weight.

### 2.4. Lesion segmentation

#### 2.4.1. Manual segmentation

For cohort 1, both 100% and 50%+PETfectior images were anonymized and sent to a GE AW Server (GE Healthcare) where four experienced physicians manually labelled every oncologic lesion by drawing a box that fully contains it as shown in Figure 2. This workstation provided threshold-based segmentation as 42% of the maximum uptake inside the bounding box. A DICOM file with the information of the lesions (position, extension, volume and SUV statistics) was generated from the saved state and exported for further analysis with custom Python code.

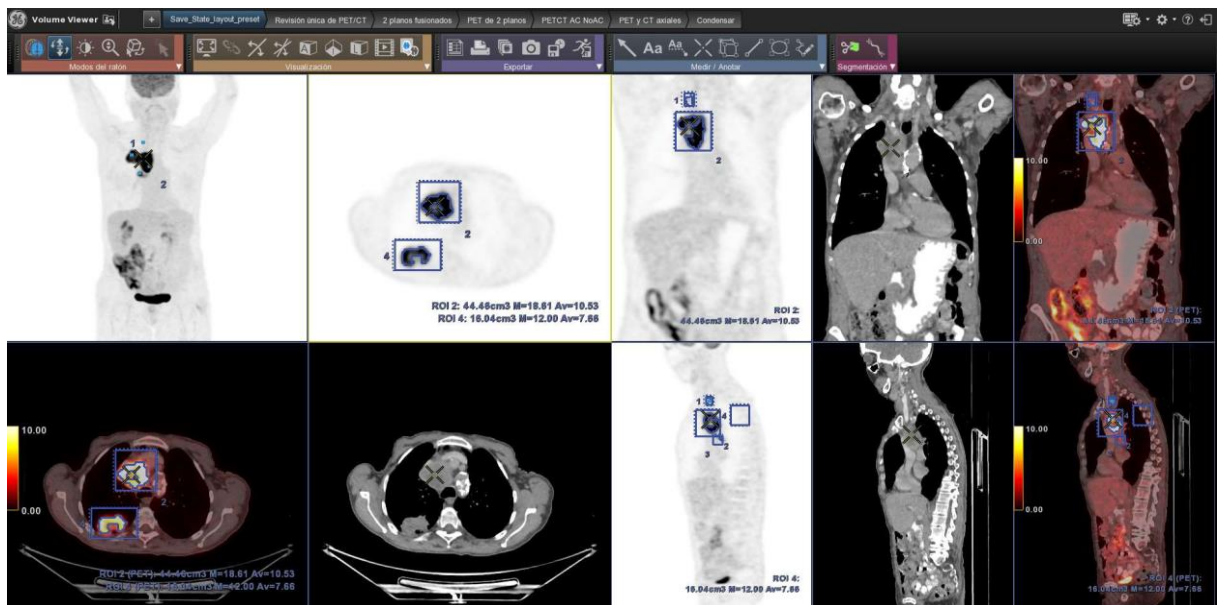

**Figure 2:** Example of the manual segmentation procedure on the AW Server (General Electric).

#### 2.4.2. Automatic segmentation

Cohort 2 was processed using an already trained convolutional neural network (Rotstein Habarnau & Namías, 2023) which was designed to participate in the AutoPET challenge (*AutoPET-II*, 2023) to automatically segment all the oncologic lesions on the PET/CT images (in both 100% and 50%+PETfectior versions). The segmentation tool provides a mask of each lesion that can be applied to the PET/CT image in order to obtain the desired information (SUV statistics, volume, etc.). The segmentation masks were validated by the physicians, greatly accelerating the procedure compared to cohort 1. This network was trained prioritizing sensitivity over specificity, so some false positives (FP) occurred. Thus, FP findings from the segmentation network (e.g.: infiltration at the injection site, physiological uptake, etc.) were visually identified and discarded from the dataset.

## 2.5. Lesion detectability and quantitative performance evaluation

### 2.5.1. Automatic segmentation

We automatically compared the masks of the lesions detected in 100% image and 50%+PETfectior image to obtain a preliminary list of true positives (TP), false negatives (FN), and false positives (FP). We considered as TP those lesions detected in the 100% image that overlapped by at least one voxel with any lesion detected in the 50%+PETfectior image. Those lesions detected in the 100% image that had no overlap with the lesions in the 50%+PETfectior image were considered FN. The lesions detected in the 50%+PETfectior image with no overlap with any lesion detected in the 100% image, were considered FP.

After this initial step, each suspected FN and FP was reviewed. To that end, the 100% and 50%+PETfectior images were visually compared in those regions where the network detected a discrepancy. If a lesion was only detected by the network in the 100% image but was clearly visible for the experts in the 50%+PETfectior image, then it was not considered as FN and was included in the dataset for quantitative assessment. We proceeded similarly with the FP, where the experts reviewed these findings to classify them as pathological or otherwise.

### 2.5.2. Manual segmentation

The DICOM files obtained after manual delineation of the lesions were processed in order to compare the lesions segmented on the 100% and 50%+PETfectior images. To that end, the DICE coefficient was used. Lesions were considered TP if they had at least a 10% overlap between 100% and 50%+PETfectior versions. Those lesions detected in the 100% image but not in the 50%+PETfectior image are considered FN. The lesions detected in the 50%+PETfectior image but not in the 100% image are considered FP.

## 2.6. Sensitivity and false positive rate evaluation

We compared the masks of the lesions detected in 100% and 50%+PETfectior images. We evaluated the number of TP, FN and FP. Besides, we computed the sensitivity of the 50%+PETfectior images as

$$S = \frac{TP}{TP + FN}.$$

## 2.7. Quantitative concordance

We compared the maximum SUV ($SUV_{max}$) measured in the 100% and 50%+PETfectior images, taking into account only those quantifiable lesions that fulfilled the following criteria (Kinahan et al., 2020):

- $SUV_{max}$ > 4 g/ml (measured in 100% image)
- volume > 4.2 cm$^3$

Agreement was assessed using Bland-Altman plots, the 95% limits of agreement (LoA), bias and mean absolute error (MAE).

## 2.8. Subjective image quality

Image quality was visually assessed by two experienced physicians (for both 100% and 50%+PETfectior versions) on cohort 1 (blind test). They were asked to rate, on a scale from 1 to 5, the characteristics detailed in Table 3.

| Characteristic | Rate |
|---|---|
| Global image quality | 1. Non-diagnostic<br>2. Poor<br>3. Acceptable<br>4. Good<br>5. Excellent |
| Global diagnostic confidence | 1. None<br>2. Poor<br>3. Acceptable<br>4. Good<br>5. Excellent |
| Image artifacts | 1. Extreme artifacts that make it impossible to interpret the images<br>2. The artifacts make the interpretation of the image difficult<br>3. Some artifacts are present, but images can be interpreted<br>4. Mild artifacts that do not affect the interpretation of the image<br>5. No visible artifacts |
| Image noise | 1. Extremely high noise level<br>2. High noise level, doubts about possible lesions<br>3. Acceptable noise level, but it could be better<br>4. Low noise level<br>5. Really low noise level, it does not interfere at all |
| Structure definition | 1. Undefined structures<br>2. Structures with poor definition<br>3. Structures with acceptable definition, it could be better<br>4. Well defined structures<br>5. Perfectly defined structures |

**Table 3:** Subjective image characteristics used to evaluate image quality.

Traditional image quality metrics such as PSNR and SSIM were not evaluated since PETfectior can provide images with higher SNR than the 100%-counting-statistics version of the images.

## 3. Results

We show in Figure 3, Figure 4, Figure 5 and Figure 6 examples of image pairs (100% and 50%+PETfectior). For one of them (Figure 3), the oncologic lesions detected by the network are shown in red.

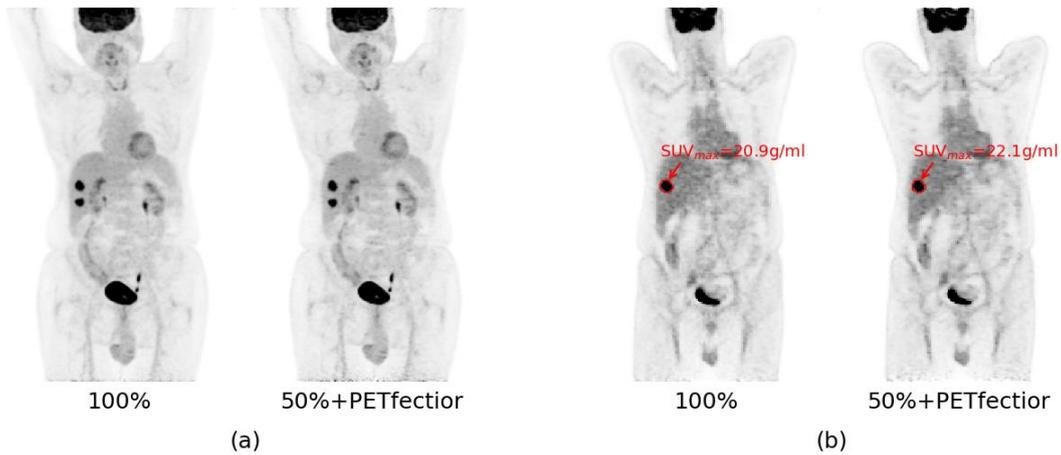

**Figure 3:** Male patient, 77 years old, 84 kg, diagnosed with colon cancer. Injected activity: 330.5 MBq. Scanner: GE Discovery 610. Maximum intensity projection (MIP) view (a) and coronal view (b) of the 100% image and the 50%+PETfectior image.

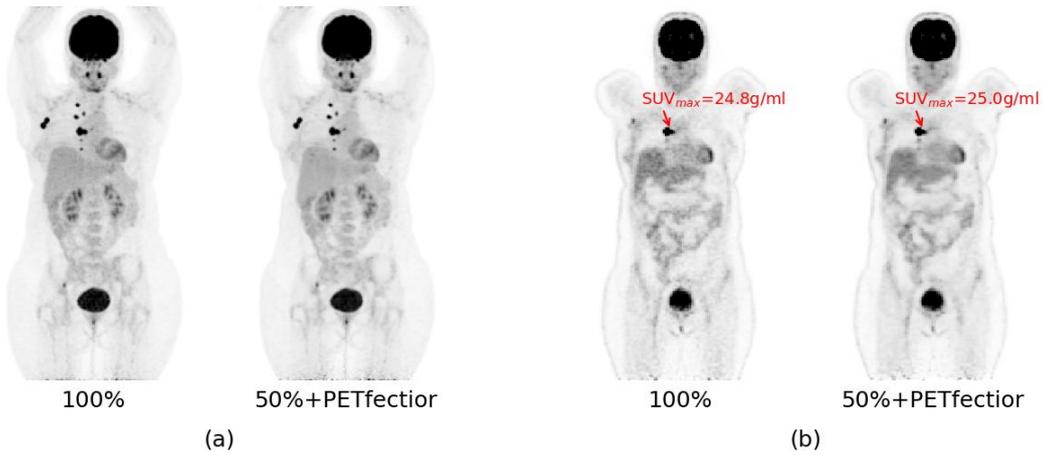

**Figure 4:** Female patient, 35 years old, 65 kg, diagnosed with breast cancer. Activity: 194.5 MBq. Scanner: GE Discovery 610. Maximum intensity projection (MIP) view (a) and coronal view (b) of the 100% image and the 50%+PETfectior image.

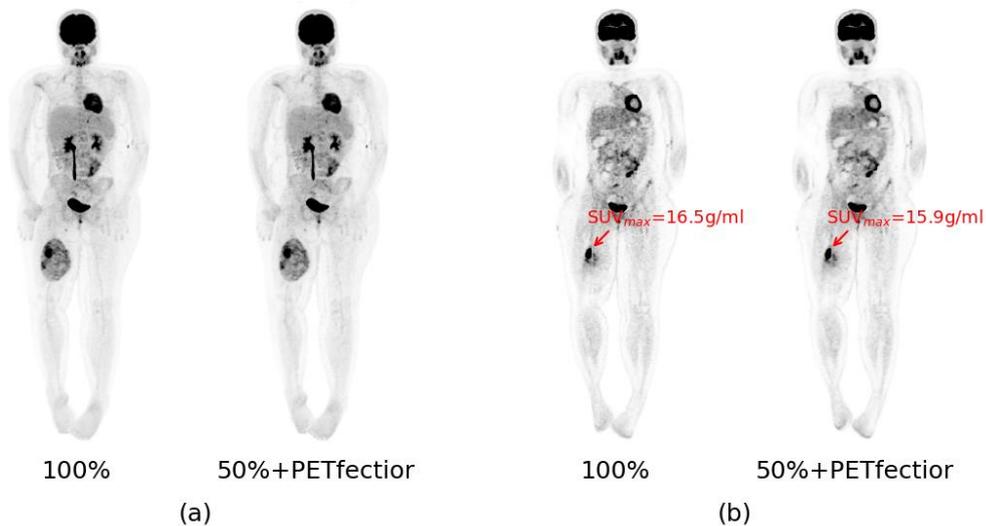

**Figure 5:** Female patient, 12 years old, 87 kg, diagnosed with rhabdomyosarcoma. Activity: 308.7 MBq. Scanner: GE Discovery 710. Maximum intensity projection (MIP) view (a) and coronal view (b) of the 100% image and the 50%+PETfectior image.

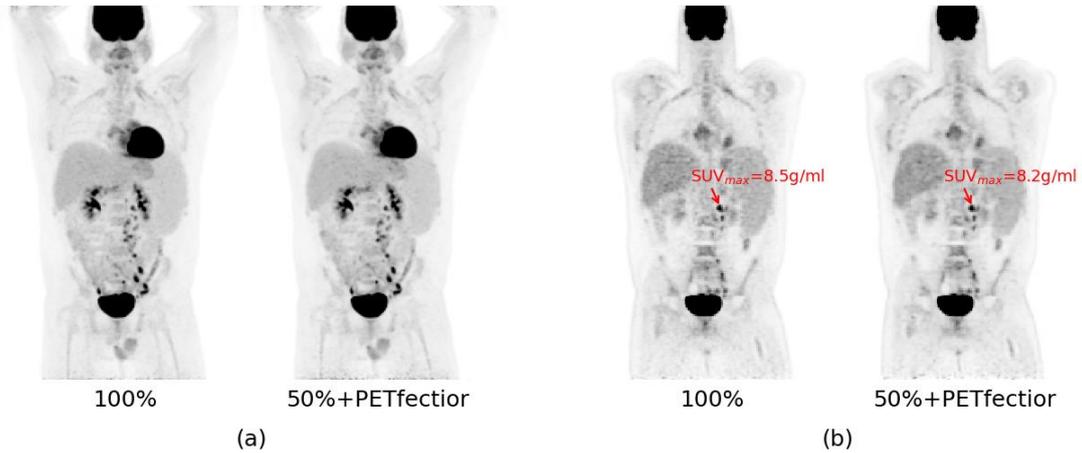

**Figure 6:** Male patient, 30 years old, 96 kg, diagnosed with Hodgkin lymphoma. Activity: 376.5 MBq. Scanner: GE Discovery 710. Maximum intensity projection (MIP) view (a) and coronal view (b) of the 100% image and the 50%+PETfectior image.

### 3.1. Lesion detectability

1649 lesions were detected in a total of 198 studies. A detail of the number of lesions per scanner can be found in Table 4. We show in Figure 7 the distribution of lesion volume and $SUV_{max}$ statistics measured in the reference image (100%).

| Scanner | Segmentation | No. of studies with lesions / total | Total no. of lesions | No. of lesions per study | | | |
|---|---|---|---|---|---|---|---|
| | | | | Min | Max | Mean | Median |
| GE Discovery 610 | Automatic | 146/203 | 1252 | 0 | 124 | 6.17 | 2.00 |
| GE Discovery 710 | Automatic | 6/9 | 30 | 0 | 20 | 3.33 | 2.00 |
| | Manual | 46/46 | 367 | 1 | 26 | 7.98 | 5.50 |

**Table 4:** Number of lesions segmented both automatic and manually.

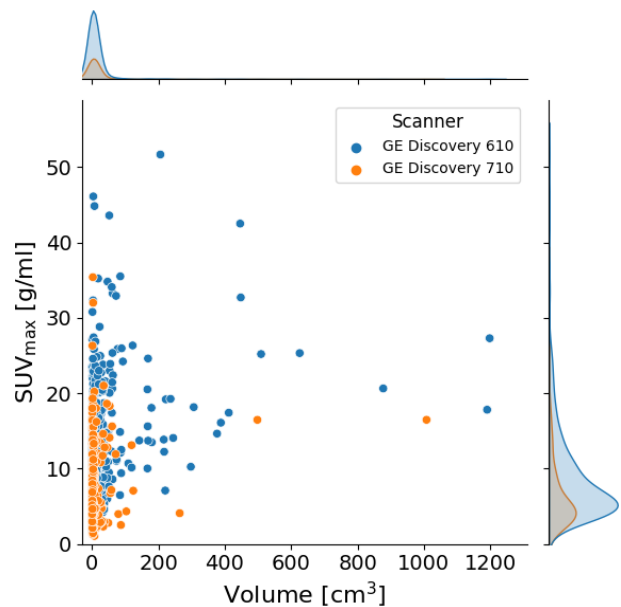

**Figure 7:** Volume and $SUV_{max}$ distribution of all the lesions detected in the reference image (100%).

### 3.1.1. Lesion detectability: GE Discovery 610

We found no false negatives, getting a sensitivity of 100%. We found 1 false positive (Figure 8). For this case, in the 50%+PETfectior image there is a high-uptake region detected as a lesion that is not visible in the 100% image and has no CT correspondence. However, as can be seen in Figure 8, this apparent lesion is already visible in the 50%-count-statistic image, prior to being processed by PETfectior and therefore it is not an artifact introduced by the method.

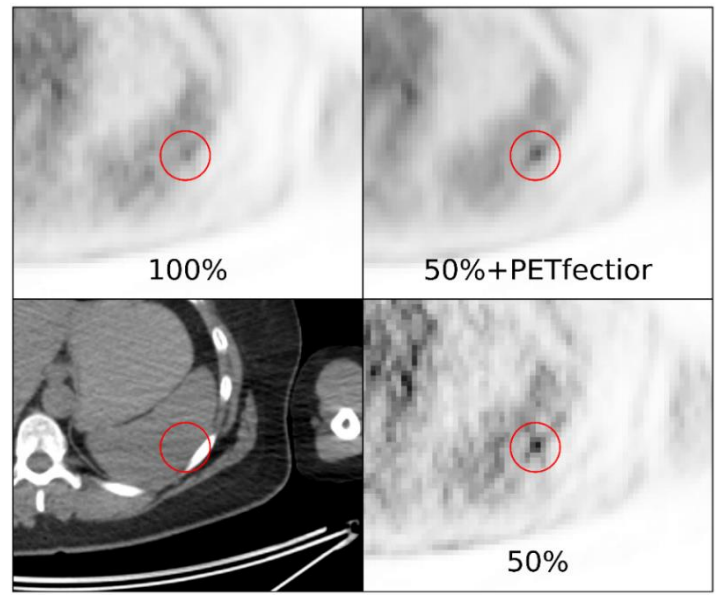

**Figure 8:** False positive finding. Female patient, 46 years old, 93 kg. Upper left: 100% image. Upper right: 50%+PETfectior version. Lower left: CT image. Lower right: 50%-count-statistic image (prior to PETfectior processing). In red, the region of the false positive finding.

Six lesions were detected in the 50%+PETfectior image but not in the 100% image. According to the experts' opinion, 4 of these segmentations did not correspond to false positives. Instead, these were real lesions that were not detected in the 100% images due to patient's motion during the longer acquisitions. We show one of these cases in Figure 9, where it is evident that it corresponds to an osteolytic lesion with CT correspondence. Two findings in the 50%+PETfectior images were inconclusive in the clinical context of the patients. In these cases, the physicians would have indicated further imaging such as delayed PET scans or MRI scans (these were not false positives that would be reported as lesions unequivocally).

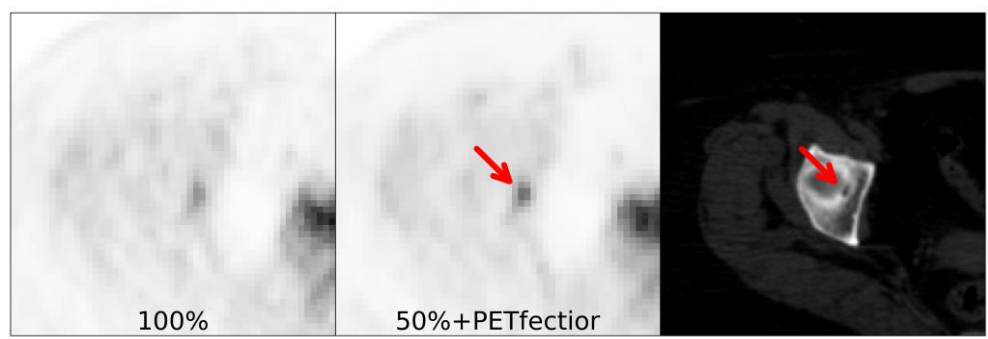

**Figure 9:** Female patient, 42 years old, 76 kg. 100% (left), 50%+PETfectior (center) and CT (right) images. The arrow points at an osteolytic lesion with CT correspondence only visible on the 50%+PETfectior version.

### 3.1.2. Lesion detectability: GE Discovery 710

408 out of the 409 lesions detected in the 100% images were also detected in their corresponding 50%+PETfectior versions, resulting in a sensitivity of 99,8%. We show in Figure 10 the false negative case. We did not find any false positives in this subset of scans.

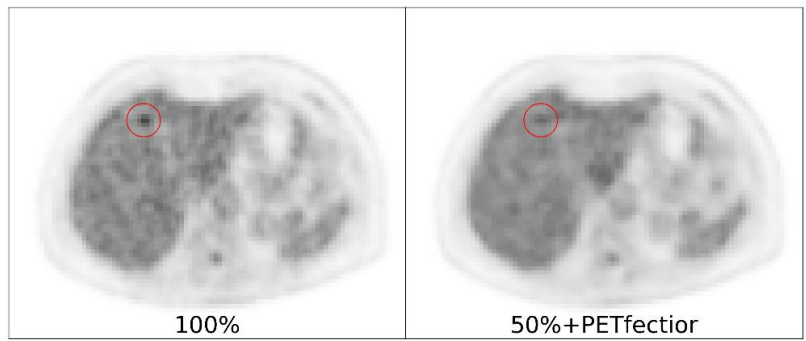

**Figure 10:** Female patient, 60 years old, 53 kg, diagnosed with colon cancer. 100% (left) and 50%+PETfectior (right) images. In red, the region of the lesion, only seen in the 100% (false negative).

### 3.2. Quantitative accuracy and reproducibility

Figure 11 shows the Bland-Altman plot demonstrating the differences in $SUV_{max}$ measured in the 100% and 50%+PETfectior images for all scanners. Figure 12 shows the Bland-Altman plot for the relative percentage differences in $SUV_{max}$ measured in both images for all scanners. The mean absolute error (MAE) of $SUV_{max}$ was 0.54 g/ml and the 95% limits of agreement (LoA) were [-1.56, 1.38] g/ml with a bias of -0.09 g/ml. As for the relative percentage differences, the 95% LoA were [-13.51, 11.48]% with a bias of -1.01% and the MAE 4.90%.

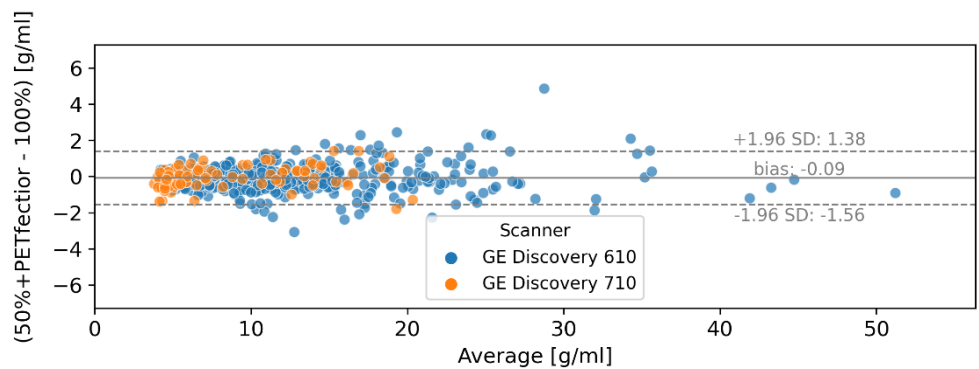

**Figure 11:** Bland-Altman plot showing differences in $SUV_{max}$ versus average $SUV_{max}$ for all scanners.

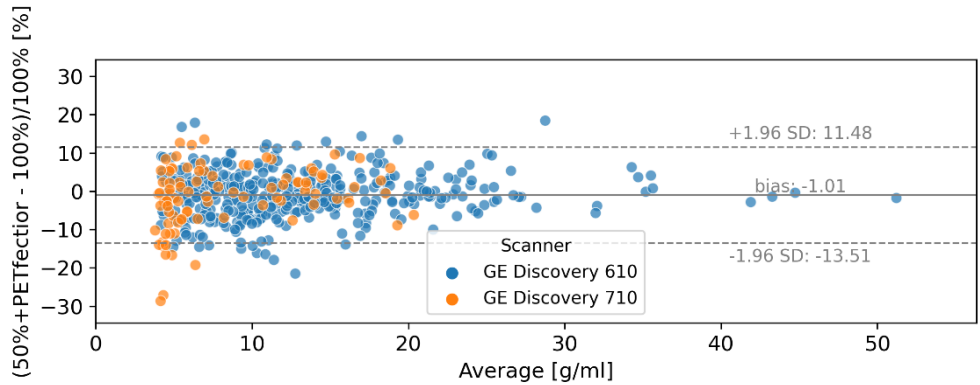

**Figure 12:** Bland-Altman plot showing relative percentage differences in $SUV_{max}$ versus average $SUV_{max}$ for all scanners.

The Bland-Altman plots for each individual scanner are shown in Figure 13 (GE Discovery 610) and Figure 14 (GE Discovery 710). The mean absolute errors, 95% limits of agreement and bias of absolute and relative $SUV_{max}$ differences for all scanners can be found in Table 5.

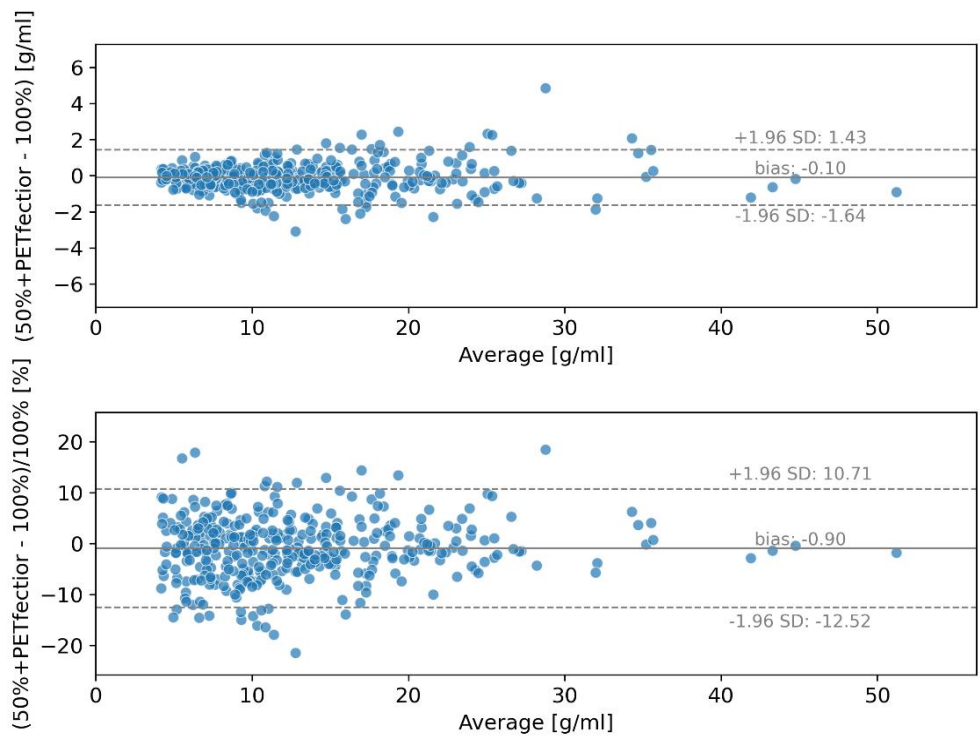

**Figure 13:** Bland-Altman plot showing absolute (upper) and relative (lower) differences in $SUV_{max}$ for GE Discovery 610 scanner.

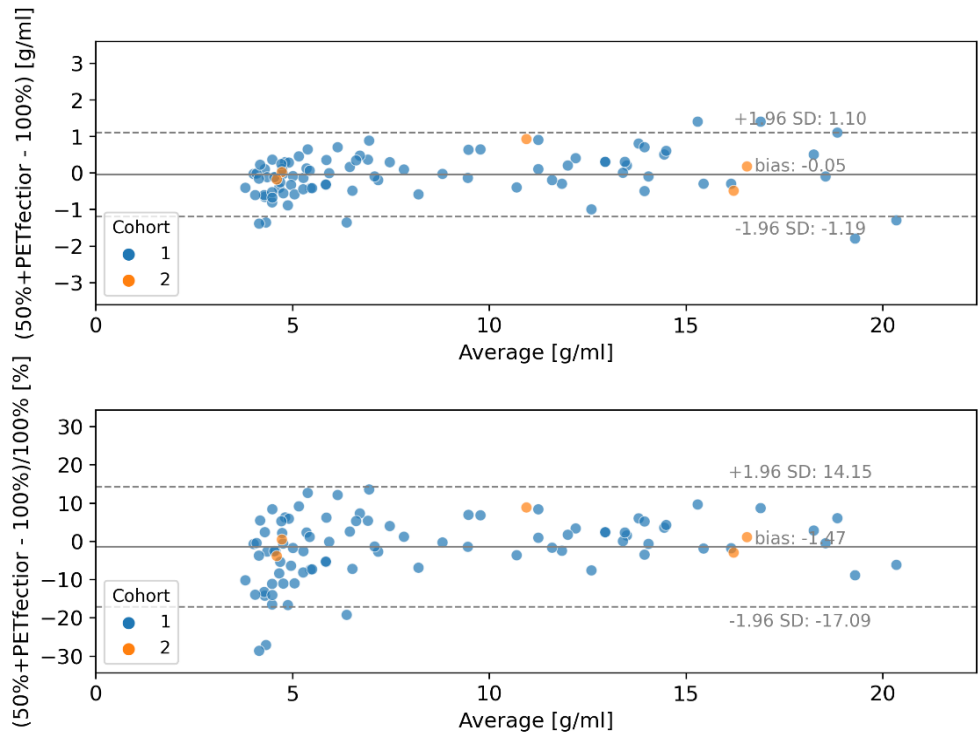

**Figure 14:** Bland-Altman plot showing absolute (upper) and relative (lower) differences in $SUV_{max}$ for GE Discovery 710 scanner.

| Scanner | $SUV_{max}$ differences [g/ml] | | | $SUV_{max}$ relative differences [%] | | |
|---|---|---|---|---|---|---|
| | Bias | 95% LoA | MAE | Bias | 95% LoA | MAE |
| GE Discovery 610 | -0.10 | [-1.64, 1.43] | 0.57 | -0.90 | [-12.52, 10.71] | 4.62 |
| GE Discovery 710 | -0.05 | [-1.19, 1.10] | 0.45 | -1.47 | [-17.09, 14.15] | 6.04 |

**Table 5:** Bias, 95% limits of agreement (LoA) and mean absolute errors (MAE) for the absolute and relative $SUV_{max}$ differences for all scanners.

### 3.3. Subjective image quality

Subjective scores assigned to 50%+PETfectior images were as good as or better than that of 100% images, as shown in Figure 15. This means that physicians rated the 50%+PETfectior images with equal or higher scores than the 100% counterparts.

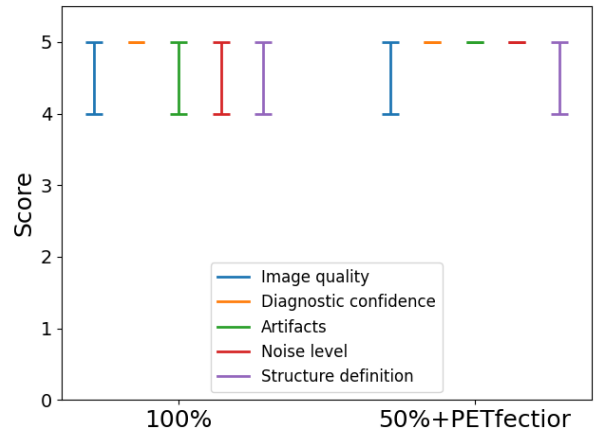

**Figure 15:** Subjective image quality scores assigned to 100% and 50%+PETfectior images.

## 4. Discussion

A total of 258 $^{18}$F-FDG PET/CT oncological studies were analyzed, comparing standard-of-care images vs half-counting-statistics versions processed by PETfectior. 1649 lesions were detected in a total of 198 studies. Unlike other validation reports found in the literature, we decided to analyze all the visible lesions and not just the main ones. As can be seen in Figure 7, the lesions present a variety of $SUV_{max}$ and volumes, although most of them are relatively small with low to moderate activity, which makes the segmentation task even more difficult. We included in our analysis all types of cancer imaged with $^{18}$F-FDG at our home institution.

As can be seen in Table 4, all the studies included in cohort 1 had lesions, with a median value of more than five lesions per study. On the other hand, cohort 2 also included some patients whose PET/CT images showed no lesions but allowed us to test whether the method introduced false positives. Thanks to the automatic lesion segmentation tool, we were able to include in cohort 2 studies with more than 100 lesions per patient.

The 50%+PETfectior images showed high sensitivity for lesion detection (99.9%) when compared with the 100% version, since there was only one false negative (Figure 10) detected on a patient scan from the GE Discovery 710 PET/CT scanner. This false negative corresponds to a lesion on the liver of a female patient diagnosed with colon cancer, with a total of 5 lesions. We only found one false positive (GE Discovery 610 scanner), located on the spleen. This study was acquired with arms down, which could lead to artifacts in the abdominal region, where the lesion was located. When analyzing the images corresponding to this false positive (Figure 8), one can notice that the FDG-avid region detected as a lesion is already visible on the half-counting-statistics image and therefore it is not an artifact introduced by PETfectior. However, given that it did not have anatomical correspondence on the CT image, it is possible that this FP is an outlier of the underlying noise distribution of the reduced-counts PET image which was not adequately suppressed by the artificial intelligence model. We also detected two findings in the liver with no CT correlation that would have required complementary studies and thus were not classified as false positives. Four actual lesions were only detected in the 50%+PETfectior image, such as the example shown in Figure 9. We attribute this phenomenon to patient's motion during longer acquisitions, demonstrating one of the advantages of being able to reduce to half the acquisition time thanks to PETfectior.

Quantitative concordance of $SUV_{max}$ was within ±12.5% (95% limits of agreement), with a bias of -1.01%. This represents a major difference compared to another commercial software, with higher reported bias and wider LoA (Bonardel et al., 2022; Weyts et al., 2022).

Regarding subjective image quality, 50%+PETfectior images had equal or higher scores than the standard-of-care images, implying that a transparent denoising process was perceived by the experts. This is important since images with a synthetic appearance might not be acceptable for clinical practice due to a lower physician confidence in the perceived image quality.

It is important to mention that both PET/CT scanners considered in this analysis were harmonized according to the latest EANM clinical standards. We did not expect major differences on the behavior of both scanners, so it is safe to assume that the appearance of one false positive on GE Discovery 610 scanner and one false negative on GE Discovery 710 scanner is not directly attributable to the scanner model. Also, the time-activity product scheme for the 100% image was already optimized as the minimum possible to comply with the EANM/EARL standards (Koopman et al., 2016). Therefore, we achieved a 50% reduction in counting statistics against an already optimized time-activity scheme. As a comparison, other commercial solutions claim a 75% dose reduction without any information loss. However, as discussed by Bonardel et al. (2022), the reference dose baseline for this method was approximately twice the recommended by the European guidelines. They found that this particular algorithm could only be used for half-time or half-dose acquisitions based on European recommended injected dose. Chaudhari et al. (2021) report good performance of this implementation on 25% low-count images, but the reported injected activity is roughly two times higher than our optimized protocols.

The validation of PETfectior for other uses, such as image quality enhancement, or other tracers was outside the scope of this article. This initial clinical validation was carried out in our home institution, and it will be extended to a multicentric study.

## 5. Conclusions

A 50% reduction in the minimum required time-activity product to comply with EANM/EARL $^{18}$F-FDG PET standards was feasible with PETfectior, with high sensitivity and low false positives per patient and high $SUV_{max}$ accuracy. Subjective image quality was as good as or better than that of the standard-of-care images. We conclude that PETfectior can safely be used in clinical practice at reduced counting statistics, with high sensitivity and specificity, low quantitative bias and high subjective image quality.